\documentclass[12pt]{article}
\usepackage{epsfig}

\usepackage{a4}

\usepackage{amsmath}

\begin{document}

\title{\bf Some aspects of Skyrme--Chern-Simons densities}
\author{ {\large D. H. Tchrakian}$^{\dagger \star}$ 
\\ 
\\
$^{\dagger}${\small School of Theoretical Physics, Dublin Institute for Advanced Studies,}
\\
{\small10 Burlington Road, Dublin 4, Ireland}
\\   
$^{\star}${\small Department of Computer Science, National University of Ireland Maynooth, Maynooth, Ireland }
}

\date{}
\newcommand{\dd}{\mbox{d}}
\newcommand{\tr}{\mbox{tr}}
\newcommand{\la}{\lambda} 
\newcommand{\La}{\Lambda}
\newcommand{\ka}{\kappa}
\newcommand{\f}{\phi}
\newcommand{\F}{\Phi}
\newcommand{\vf}{\varphi}
\newcommand{\ta}{\theta}
\newcommand{\al}{\alpha}
\newcommand{\bt}{\beta}
\newcommand{\ga}{\gamma}
\newcommand{\de}{\delta}
\newcommand{\si}{\sigma}
\newcommand{\Si}{\Sigma}
\newcommand{\bnabla}{\mbox{\boldmath $\nabla$}}
\newcommand{\bomega}{\mbox{\boldmath $\omega$}}
\newcommand{\bOmega}{\mbox{\boldmath $\Omega$}}
\newcommand{\bsi}{\mbox{\boldmath $\sigma$}}
\newcommand{\bchi}{\mbox{\boldmath $\chi$}}
\newcommand{\bal}{\mbox{\boldmath $\alpha$}}
\newcommand{\bpsi}{\mbox{\boldmath $\psi$}}
\newcommand{\brho}{\mbox{\boldmath $\varrho$}}
\newcommand{\beps}{\mbox{\boldmath $\varepsilon$}}
\newcommand{\bxi}{\mbox{\boldmath $\xi$}}
\newcommand{\bbeta}{\mbox{\boldmath $\beta$}}
\newcommand{\ee}{\end{equation}}
\newcommand{\eea}{\end{eqnarray}}
\newcommand{\be}{\begin{equation}}
\newcommand{\bea}{\begin{eqnarray}}

\newcommand{\ii}{\mbox{i}}
\newcommand{\e}{\mbox{e}}
\newcommand{\pa}{\partial}
\newcommand{\Om}{\Omega}
\newcommand{\vep}{\varepsilon}
\newcommand{\vr}{\varrho}
\newcommand{\bfph}{{\bf \phi}}
\newcommand{\om}{\omega}
\def\theequation{\arabic{equation}}
\renewcommand{\thefootnote}{\fnsymbol{footnote}}
\newcommand{\re}[1]{(\ref{#1})}
\newcommand{\R}{{\rm I \hspace{-0.52ex} R}}
\newcommand{\N}{{\sf N\hspace*{-1.0ex}\rule{0.15ex}%
{1.3ex}\hspace*{1.0ex}}}
\newcommand{\Q}{{\sf Q\hspace*{-1.1ex}\rule{0.15ex}%
{1.5ex}\hspace*{1.1ex}}}
\newcommand{\C}{{\sf C\hspace*{-0.9ex}\rule{0.15ex}%
{1.3ex}\hspace*{0.9ex}}}
\newcommand{\eins}{1\hspace{-0.56ex}{\rm I}}
\renewcommand{\thefootnote}{\arabic{footnote}}

\def\theequation{\thesection.\arabic{equation}}

\maketitle


\bigskip

\begin{abstract}
The gauge transformation properties of the Skyrme--Chern-Simons (SCS) densities is studied. Two types of SCS actions
are identified, Type$_{I}$ in which the gauge group is smaller than the largest possible one, and
Type$_{II}$ which are gauged with the largest allowed gauge group. Type$_{I}$ SCS feature only one power of the
gauge connection and no curvature, while Type$_{II}$ feature both the gauge connection and the curvature. The Abelian Type$_{I}$
SCS turn out to be explicitly gauge invariant while non-Abelian Type$_{I}$ and all Type$_{II}$ SCS are gauge invariant
only up to a total divergence term, and hence lead to gauge covariant equations of motion. SCS actions are the
gauged Skyrmion analogues of the usual Chern-Simons (CS) actions, except that unlike the CS which are defined only in odd
dimensions, the SCS are defined also in even dimensions.  Some areas of application in the construction of
solitons are pointed out.
\end{abstract}
\medskip
\medskip

\section{Introduction}
The role of the Chern-Simons (CS) action in soliton physics was recognised a long time ago in the construction of
electrically-charged spinning vortices of $SO(2)$ gauged Higgs model~\cite{Paul:1986ix,Hong:1990yh,Jackiw:1990aw} and $O(3)$
Skyrme model~\cite{Ghosh:1995ze,Kimm:1995mi,Arthur:1996uu} in $2+1$ dimensions, and,  $SO(2)$ gauged $O(5)$
Skyrme model~\cite{Navarro-Lerida:2020jft} in $4+1$ dimensions. In the case of the gauged Skyrme
solitons~\cite{Arthur:1996uu,Navarro-Lerida:2020jft}, an energy lower bound departing from the topological
``baryon number'' due to the gauge field was applied. This new ``deformed baryon number'' was first proposed in
\cite{Tchrakian:1997sj} and applied many times since, and most recently is elaborated
in Appendix {\bf B} of \cite{Tchrakian:2015pka}. The gauging prescription for the $O(D+1)$ Skyrme scalar on $\R^D$ proposed in
\cite{Tchrakian:2015pka} (and references therein) is effected by the $SO(N)\,,\ (2\le N\le D)$ connection gauging $N$
components of the $D+1$ component Skyrme scalar.

Further to the construction of solitons of Abelian gauged Skyrme models in $2+1$ and $4+1$ dimensions, remarkable
dynamical effects resulting from the Chern-Simons action were discovered in $2+1$ dimensions in
\cite{Navarro-Lerida:2016omj,Navarro-Lerida:2018siw,Navarro-Lerida:2018giv}, and in $4+1$ dimensions, in
Ref.~\cite{Navarro-Lerida:2020hph}
fields encoding the CS action in a specific way. They are $a)$ the non-standard dependence of the mass/energy $E$
on the electric charge $Q_e$ and the angular momentum $J$, whose slope can be negative (as well as positive) in
some regions of the parameter space, and $b)$ the evolution of the ``topological charge'' (baryon number) away
from its integer value prior to gauging, due to its deformation resulting from the gauging.

Thus in 'all odd dimensions' where the Chern-Simons density is defined,
namely through the examples of $2+1$ and $4+1$ cited in the previous paragraph, we
learn that the $E\ vs.\ Q_e$ and $E\ vs.\ J $ slopes can be negative as a result of the Chern-Simons dynamics.
Now it is a longstanding problem to explain why the electrically neutral Neutron described by the Skyrmion is
heavier than the electrically charged Proton which is presumably described by the Abelian gauged Skyrmion.
The latter should display a negative $E\ vs.\ Q_e$ slope. But in $3+1$ dimensions there is no Chern-Simons term
defined, so the Skyrmion of the Abelian gauged $O(4)$ sigma model cannot be influenced by this mechanism
to produce a negative $E\ vs.\ Q_e$ slope. It would be desirable for this purpose to avail of a
Chern-Simons like density in even dimensions. Such an ``anomaly associated'' density with the appropriate parity
reversal properties, defined in terms of the
Abelian field interacting with the Skyrme scalar in $3+1$ dimensions, was displyed
in Ref.~\cite{Callan:1983nx},  albeit not in the context of the $E\ vs.\ Q_e$ slope question.
The SCS actions in $3+1$ dimensions can be seen as a possible alternative to the ``anomaly associated''
action of Ref.~\cite{Callan:1983nx}.

It is precisely to fill this gap that the Skyrme--Chern-Simons (SCS) densities, which are defined in both even and
odd dimensions, were proposed in Ref.~\cite{Tchrakian:2015pka}.
The present note is intended to clarify and elucidate aspects of these SCS densities, principally to demonstrate
that the variational equations of these actions are {\it gauge covariant}~\footnote{
It could be mentioned in passing that in addition to the SCS actions, the so-called Higgs---Chern-Simons (HCS) densities
are proposed in \cite{Tchrakian:2015pka}, lead to {\it gauge covariant} variational equations. This aspect is obvious
in the case of HCS densities, since in even dimensions these are manifestly {\it gauge invariant} and in odd
dimensions they consist of a Higgs dependant part which is gauge invariant, plus, the usual Chern-Simons density
in the given (odd) dimension.}. In this note two types of SCS densities are identified, Type$_{I}$ and Type$_{II}$.
Type$_{I}$ SCS densities are those which are gauged with the single gauge group $SO(N)$, where $N<d+1$, $SO(d+1)$
being the largest gauge group a SCS density in $d$ dimensions can be gauged with. The typical feature of Type$_{I}$
is that only a single connection $A_\mu$ appears and no curvature $F_{\mu\nu}$ is displayed. A technical feature of
Type$_{I}$ SCS is that they can be expressed in closed form in all dimensions $d$, for each $SO(N)$.
Type$_{II}$ SCS densities on the other hand are those pertaining to the largest allowed gauge group in $d$
dimensions, $SO(d+1)$, as well as those gauged with the direct product of all possible subgroups $SO(d+1)$.
The important distinction between Type$_{I}$ and Type$_{II}$ SCS is that the former display one power of the
connection $A_\mu$ only, while the later display both connection and curvature
$(A_\mu,F_{\mu\nu})$ for $SO(d+1)$, or as the case may be, for each of the (direct product) subgroups therein.
In this sense Type$_{II}$ SCS are the germane analogues of the usual Chern-Simons (CS) terms which are expressed in terms
of $(A_\mu,F_{\mu\nu})$, except that unlike the CS
which defined in only odd $d$, the Type$_{II}$ SCS are defined in all $d$.

In Section {\bf 2}, a revision of the definition of the usual Chern-Simons (CS) densities is given, followed by a
brief presentation of the generic Skyrme--Chern-Simons (SCS) densities. The Type$_{I}$
SCS densities in arbitrary $d$ dimensions for gauge groups $SO(2)$ and $SO(3)$ are presented in Section {\bf 3},
one Abelian and one non-Abelian example, the smallest of $SO(N)\ \ (N<d+1)$ gauge groups.
Section {\bf 4} deals in principle with Type$_{II}$  $SO(d+1)$ (and all available direct product subgroups of)
SCS in $d$ dimensions, but in practice it is restricted to
dimension $d=3$ with gauge group $SO(4)$ and its direct product subgroup $SO(2)\times SO(2)$.
Section {\bf 5} gives a summary and some possible application to solitons are pointed out.

\section{Chern-Simons (CS) and Skyrme--Chern-Simons (SCS)}
In the first Subsection, a brief review of the usual Chern-Simons densities is presented with the purpose of
putting into context the definition of the Skyrme--Chern-Simons densities proposed in \cite{Tchrakian:2015pka},
which is summarised in the subsequent Subsection below.

\subsection{Brief review of Chern-Simons (CS)}
The starting point in the definition of the CS density in $d$ dimensions is the Chern-Pontryagin (CP) density
in $d+1$, $even$, dimensions
\be
\label{CP0}
\vr\equiv\Om^{(d+1)}_{\rm CP}=\vep_{i_1i_2i_3i_4\dots i_d i_{d+1}}\,\mbox{Tr}\,F_{i_1i_2}F_{i_3i_4}\dots F_{i_d i_{d+1}}\,,
\ee
which happens to be the topological charge density stabilising $SO_{\pm}(d+1)$ ``instantons'' in all even dimensions
(see $e.g.,$ \cite{Tchrakian:2010ar}). The density \re{CP0} is known to be a total divergence.

The usual Chern-Simons (CS) density $\Om^{(d)}_{\rm CS}$ in $d$ ({\it odd}) dimensions
is extracted from the ({\it total divergence})
Chern-Pontryagin (CP) density in $d+1$ ({\it even}) dimensions
\be
\label{CP}
\Om^{(d+1)}_{\rm CP}=\pa_i\Om^{(d+1)}_i\,\ , \quad i=1,2,\dots,d+1 
\ee
as the $(d+1)$-th component of $\Om^{(d+1)}_i$ in \re{CP},
\be
\label{CS}
\Om^{(d)}_{\rm CS}\stackrel{\rm def.}=\Om^{(d+1)}_{i=d+1}\,,
\ee
and since $\Om^{(d+1)}_{\rm CP}$ is a ``curl'' defined in terms of the totally antisymmetric tensor
$\vep^{i_1i_2\dots i_{d+1}}$, fixing one component, say $i=d+1$, is tantamount to a descent by one
dimension, such that $\Om^{(d)}_{\rm CS}$ defined by \re{CS} is a scalar in $d$ dimensions.

Thus, the CS density \re{CS} expressed in terms of the gauge connection $A_\mu$ and the
curvature $F_{\mu\nu}$ is defined in a $d$ dimensional space with coordinates $x_\mu$
\be
\label{AF}
\Om^{(d)}_{\rm CS}=\Om^{(d)}_{\rm CS}[A_\mu,F_{\mu\nu}] ,\quad \mu=1,2,\dots,d\ ,\quad d\ \ {\rm odd}\,.
\ee

Most remarkably, CS densities and are {\it explicitly gauge variant}, as implied by the notation used in \re{AF}.
It is important to stress that theories endowed with dynamical Chern-Simons (CS) terms in the Lagrangian are
defined on spacetimes with Minkowskian signature. Since the CS term is independent of the metric tensor, the
resulting Stress Tensor does not feature it and the static 
Hamiltonian (and hence energy) is {\it gauge invariant} as it should be~\footnote{Should one
employ a CS density on a space with Euclidean signature, with the CS density appearing in
the static Hamiltonian itself, then the energy would not be $gauge$ $invariant$. Hamiltonians of this
type have been considered in the literature, $e.g.$, in \cite{Rubakov:1986am}.

Chern-Simons densities
on Euclidean spaces, defined in terms of the composite connection of a sigma model, find application as the
topological charge densities of Hopf solitons.}.

Of course, the CP densities and the resulting CS densities, can be defined in terms of both Abelian and non-Abelian
gauge connections and curvatures. The context of the present notes is the construction of soliton
solutions~\footnote{The term soliton solutions here is used rather loosely, implying only the construction of
regular and finite energy solutions, without insisting on topological stability in general.}, rather than the study of
topologically massive field theories as in \cite{Deser:1982vy,Deser:1981wh}. In this respect,
the choice of gauge group any given dimension  must be made with due regard to
regularity, and the models chosen must be consistent with the Derrick scaling requirement for the finiteness of
energy. Accordingly, in all but $2+1$ dimensions, our considerations are restricted to non-Abelian gauge fields.

\subsubsection{Gauge transformation of CS}
Consider the transformation of $\Om^{(d+1)}_i$ in \re{CP} under the infinitesimal gauge transformation $g(x_i)$
\be
\label{infOmdp1}
\Om^{(d+1)}_i\stackrel{g}\rightarrow\Om^{(d+1)}_i+\de\Om^{(d+1)}_i\,.
\ee

Since $\vr=\pa_i\Om^{(d+1)}_i$ is gauge invariant, $i.e.,$ $\de\vr=0$, it follows that
\be
\label{pahatOm}
\pa_i(\de\Om^{(d+1)}_i)=0\,,
\ee
which allows to express $\de\Om^{(d+1)}_i$ formally as
\be
\label{Omv0}
\de\Om^{(d+1)}_i=\vep_{ijk_1k_2\dots k_{d-1}}\,\pa_jV_{k_1k_2\dots k_{d-1}}\,,
\ee
where $V_{k_1k_2\dots k_{d-1}}$ is a totally antisymmetric tensor defined in terms of the connection and the curvature fields.

From the definition \re{CS} of the CS density, \re{Omv0} implies the following transformation
\be
\label{deOmCS}
\de\Om^{(d)}_{\rm CS}=\de\Om^{(d+1)}_{i=d+1}=
\vep_{(d+1)\mu\nu_1\nu_2\dots\nu_{d-1}}\,\pa_\mu V_{\nu_1\nu_2\dots\nu_{d-1}}
\ee
which is clearly defined on the space with the $d$-dimensional coordinates $x_\mu$.

It follows from \re{deOmCS} that under an infinitesimal gauge transformation $g(x_\mu)$, the CS density
$\Om^{(d)}_{\rm CS}$ transforms as
\be
\label{infOmCS}
\Om^{(d)}_{\rm CS}\stackrel{g}\rightarrow
\Om^{(d)}_{\rm CS}+\vep_{\mu\nu_1\nu_2\dots\nu_{d-1}}\,\pa_\mu V_{\nu_1\nu_2\dots\nu_{d-1}}\,,
\ee
meaning that the CS density is {\it gauge invariant up to a total divergence}. 
One concludes that the action of $\Om^{(d)}_{\rm CS}$, namely its volume integral
\[
\int\,d^dx\ \Om^{(d)}_{\rm CS}\stackrel{g}\to\int\,d^dx\ \Om^{(d)}_{\rm CS}
\]
remains invariant under the action of $g$, resulting in the Euler-Lagrange equations \re{ELCS3} and \re{ELCS5}
being {\it gauge covariant}.

This statement can be given concrete expression in terms of examples in dimensions $d=3,5,7$, which
can be extended to all odd dimensional spacetimes systematically.

The CS densities $\Omega_{\rm CS}^{(d)}$, defined by \re{CS}, for $d=3,5,7$, are
\bea
\Omega_{\rm CS}^{(3)}&=&\vep_{\la\mu\nu}\mbox{Tr}\,
A_{\la}\left[F_{\mu\nu}-\frac23A_{\mu}A_{\nu}\right],
\label{CS3}
\\
\Omega_{\rm CS}^{(5)}&=&\vep_{\la\mu\nu\rho\si}\mbox{Tr}\,
A_{\la}\left[F_{\mu\nu}F_{\rho\si}-F_{\mu\nu}A_{\rho}A_{\si}+
\frac25A_{\mu}A_{\nu}A_{\rho}A_{\si}\right],
\label{CS5}
\\
\Omega_{\rm CS}^{(7)}&=&\vep_{\la\mu\nu\rho\si\tau\ka}
\mbox{Tr}\,A_{\la}\bigg[F_{\mu\nu}F_{\rho\si}F_{\tau\ka}
-\frac45F_{\mu\nu}F_{\rho\si}A_{\tau}A_{\ka}-\frac25
F_{\mu\nu}A_{\rho}F_{\si\tau}A_{\ka}
\nonumber
\\
&&\qquad\qquad\qquad\qquad\qquad\qquad
+\frac45F_{\mu\nu}A_{\rho}A_{\si}A_{\tau}A_{\ka}-\frac{8}{35}
A_{\mu}A_{\nu}A_{\rho}A_{\si}A_{\tau}A_{\ka}\bigg]\,.\label{CS7}
\eea

Restricting to orthogonal groups, one notes that the
CS term in $d$ dimensions features the product of $d-1$ powers of the (algebra valued) gauge field/connection in front
of the Trace, which would vanish if the gauge group $is\ not\ larger\ than$ $SO(d-1)$. In that case, the YM connection would
describe only a 'magnetic' component, with the 'electric' component necessary for the the nonvanishing of the CS
density would be absent. As in \cite{Brihaye:2009cc}, the
most convenient choice is $SO(d+1)$. Since $d+1$ is always even, the representation of $SO(d+2)$ are the $chiral$
representation in terms of (Dirac) spin matrices. This completes the definition of the usual non-Abelian Chern-Simons
densities in $d$ spacetimes.

From \re{CS3}-\re{CS7}, it is clear that the CS density is explicitly gauge variant. Their Euler-Lagrange equations
w.r.t. the variation of $A_\la$ are nonetheless gauge invariant
\bea
\delta_{A_{\la}}\Omega_{\rm CS}^{(2)}&=&\vep_{\la\mu\nu}F_{\mu\nu},\label{ELCS3}\\
\delta_{A_{\la}}\Omega_{\rm CS}^{(3)}&=&\vep_{\la\mu\nu\rho\si}F_{\mu\nu}F_{\rho\si},\label{ELCS5}\\
\delta_{A_{\la}}\Omega_{\rm CS}^{(4)}&=&\vep_{\la\mu\nu\rho\si\ka\eta}F_{\mu\nu}F_{\rho\si}F_{\ka\eta}\,.\label{ELCS7}
\eea

This remarkable property of CS densities can be understood by noting that, while these appear as explicitly
gauge-variant densities, these actions are actually {\it gauge invariant up to a surface term}. To see this one
subjects them to transformation under the action of an element, $g$ of the (non-Abelian) gauge group.

The transformations for the two examples \re{CS3} and \re{CS5} are explicitly
\bea
\Omega_{\rm CS}^{(3)}&\stackrel{g}\to&\Omega_{\rm CS}^{(3)}-
\frac23\vep_{\la\mu\nu}\mbox{Tr}\,\al_{\la}\al_{\mu}\al_{\nu}
-2\vep_{\la\mu\nu}\,\pa_{\la}\mbox{Tr}\,\al_{\mu}\,A_{\nu},\label{gaugeCS3}\\
\Omega_{\rm CS}^{(5)}&\stackrel{g}\to&\Omega_{\rm CS}^{(5)}-\frac25\,\vep_{\la\mu\nu\rho\si}
\mbox{Tr}\,\al_{\la}\al_{\mu}\al_{\nu}\al_{\rho}\al_{\si}\nonumber\\
&&+
2\,\vep_{\la\mu\nu\rho\si}\,\pa_{\la}\mbox{Tr}\,\al_{\mu}\bigg[
A_{\nu}\left(F_{\rho\si}-\frac12A_{\rho}A_{\si}\right)+\left(F_{\rho\si}-\frac12A_{\rho}A_{\si}\right)A_{\nu}\nonumber\\
&&\qquad\qquad\qquad\qquad\qquad\qquad\qquad\qquad-\frac12\,A_{\nu}\,\al_{\rho}\,A_{\si}-\al_{\nu}\,\al_{\rho}\,A_{\si}
\bigg]\,,\label{gaugeCS5}
\eea
where $\al_{\mu}=\pa_{\mu}g\,g^{-1}$, as distinct from the algebra valued quantity $\beta_{\mu}=g^{-1}\,\pa_{\mu}g$ that
appears as the inhomogeneous term in the gauge transformation of the non-Abelian connection. \re{gaugeCS3} and
\re{gaugeCS5} are the explicit versions of \re{infOmCS}.

As seen from \re{gaugeCS3}-\re{gaugeCS5}, the gauge variation of $\Omega_{\rm CS}^{(d)}$ consists of a term which is
explicitly a total divergence, and, another term
\be
\label{om}
\omega^{(d)}=\frac{2}{d}\,\vep_{\mu_1\mu_2\dots\mu_{d}}\mbox{Tr}\,\al_{\mu_1}\al_{\mu_2}\dots\al_{\mu_{d}}\,,
\ee
which in the appropriate parametrisation can be cast in the form of a winding number density.



\subsection{Brief definition of Skyrme--Chern-Simons (SCS)}
This Subsection is intended to provide a self-contained definition of SCS densities, which are discussed at length in
\cite{Tchrakian:2015pka}.

The $O(d+2)$ sigma system is defined in terms of the Skyrme scalar $\f^a\ ,\ a=1,2,\dots,d+2$, subject to $|\f^a|^2=1$.
The Skyrme--Chern-Simons (SCS)
density $\Om^{(d)}_{\rm SCS}$ in $d$ dimensions~\footnote{With solitons in mind, this is taken to be a $d$
dimensional Minkowski space, but the choice of
this signature is not important in principle.} is defined in terms of the Skyrme scalar $\f^a$.

The definition of the Skyrme--Chern-Simons (SCS) density in $d$ dimensions follows exactly analogously to the
step \re{CP}$\to$\re{CS} by the one-step descent of the Chern-Pontryagin density $\Om^{(d+1)}_{\rm CP}$ to the
Chern-Simons density $\Om^{(d+1)}_{\rm CS}$. This is done by carrying out such a descent of a density in $d+1$ dimensions
that might be denoted~\footnote{The subscript on $\Om^{(d)}_{\rm SCP}$ is purely symbolic and is meant to
underline the analogy with the usual Chern-Pontryagin density. Taking this
nomenclature literally, as Skyrme--Chern-Pontryagin, is misleading. It is the density presenting the lower bound on
the ``energy'' of the gauged Skyrme system, and is deformation of the (topological) winding number density \re{wnd}.} as
$\Om^{(d+1)}_{\rm SCP}$, down to the Skyrme--Chern-Simons (SCS) density
\bea
\Om^{(d+1)}_{\rm SCP}&=&\pa_i\hat\Om^{(d+1)}_i\label{SCP}\\
&{\rm and}&\nonumber\\
\Om^{(d)}_{\rm SCS}&\stackrel{\rm def.}=&\hat\Om^{(d+1)}_{i=d+1}\label{SCS1}\,.
\eea

Just as the CP density $\Om^{(d+1)}_{\rm CP}$ in \re{CP} is both gauge invariant and total divergence, so too is
$\Om^{(d+1)}_{\rm SCP}$ in \re{SCP}, and this is expressed symbolically in terms of the vector valued density
$\hat\Om^{(d+1)}_i$. The definitions of these quantities is given below.

The winding number density of the Skyrme scalar $\f^a\,,\ a=1,2,\dots,d+1$ of the $O(d+2)$ sigma model,
subject to the constraint $|\f^a|^2=1$ is
\be
\label{wnd}
\vr_0=\vep_{i_1i_2\dots i_{d+1}}\pa_{i_1}\f^{a_1}\pa_{i_2}\f^{a_2}\dots\pa_{i_{d+1}}\f^{a_{d+1}}\f^{a_{d+2}}
\ee
of the $O(d+2)$ Skyrme scalar $\f^a$ in $d$ dimensions, which is {\it essentially total divergence},
 and in constraint compliant parametrisation it is {\it explicitly total divergence}.
Its volume integral is a topological charge. It is not gauge invariant.

Replacing the partial derivatives $\pa_i\f^a$ in \re{wnd} with the covariant derivatives $D_i\f^a$
\be
\label{wnG}
\vr_G=\vep_{i_1i_2\dots i_{d+1}}D_{i_1}\f^{a_1}D_{i_2}\f^{a_2}\dots D_{i_{d+1}}\f^{a_{d+1}}\f^{a_{d+2}}
\ee
which is gauge invariant but not total divergence.

The density $\vr_0$ gives the lower bound for the ``energy'' of the unguaged Skyrmion, but is inadequate for
this purpose after gauging since it is not gauge invariant. The density $\vr_G$ on the other hand is gauge
invariant, but again is inadequate for this purpose as it is not a total divergence. What is needed is
a density, say $\vr\stackrel{\rm def.}=\Om^{(d+1)}_{\rm SCP}$, which is both gauge invariant and total divergence
like the CP density $\Om^{(d+1)}_{\rm CP}$ in \re{CP}. To this end,
one evaluates the difference $(\vr_G-\vr_0)$, which can be cast in the following form
\be
\label{dif}
\vr_G-\vr_0=\pa_i\Om^{(d+1)}_i[A,\f]-W[F,D\f]
\ee
where $W[F,D\f]$ is gauge invariant by construction and $\Om^{(d+1)}_i[A,\f]$ is gauge variant. In certain
examples, there is some arbitrariness in the splitting into these two terms. This will be pointed out in
Section {\bf 2.2.1} below.

The relation \re{dif} is evaluated explicitly in each $d+1$ dimension, and in each case with
the gauge group $SO(N)\ ,\ 2\le N\le d+1$, that defines the covariant derivative in \re{wnG}. The
calculations are carried out directly, using the Leibnitz rule and the tensor identities.

Collecting the gauge invariant pieces $\vr_G$ and $W$ in \re{dif}, and separately,
the individually gauge variant pieces $\vr_0$ and $\pa_i\Om^{(d+1)}_i$,
one has two equivalent definitions of a density
\bea
\vr&=&\vr_G+W[F,D\f]\label{vr1}\\
&=&\vr_0+\pa_i\Om^{(d+1)}_i[A,\f]\,,\label{vr2}
\eea
which is adopted as the definition for the density $\vr\stackrel{\rm def.}=\Om^{(d+1)}_{\rm SCP}$ presenting a lower
bound on the ``energy'' in the same way as does the usual CP density.

The two equivalent definitions \re{vr1} and \re{vr2} of $\vr$ are, as required, both gauge invariant and (essentially)
total divergence. The quantity $\vr$ is the deformation of the of $\vr_0$, namely of the (topological) ``baryon number''
when the gauge field is switched on.

Noting that $\vr_0$ is {\it essentially total divergence}~\footnote{The meaning of the phrase
{\it essentially total divergence} used here is,
that this quantity is not explicitly total divergence, but rather that the resulting Euler-Lagrange equations are trivial as
in the case of an explicitly {\it total divergence} density. Throughout, the term {\it total divergence} is used as a synonym for
{\it explicitly total divergence}.}
and hence in a constraint compliant parametrisation it is
{\it explicitly total divergence}, say
\be
\label{paom}
\vr_0=\pa_i\om_i^{(d+1)},
\ee
\re{vr2} can be expressed (explicitly) as the total divergence $\pa_i\hat\Om^{(d+1)}_i$ in \re{SCP}.

Thus, the expression \re{vr2} for $\vr$ can be written as
\bea
\vr&=&\pa_i(\om^{(d+1)}_i+\Om^{(d+1)}_i)\label{defhatOm}\\
&\stackrel{\rm def.}=&\pa_i\hat\Om^{(d+1)}_i\equiv\Om_{\rm SCP}^{(d+1)}\label{13}\,,
\eea
from which follows immediately, the definition of the SCS density \re{SCS1}
\bea
\Om^{(d)}_{\rm SCS}=\hat\Om^{(d+1)}_{i=d+1}&=&\om^{(d+1)}_{i=d+1}+\Om^{(d+1)}_{i=d+1}\label{SCS0}\\
&=&\om^{(d)}+\Om^{(d)}\label{SCSf}
\eea
where we have denoted $\om^{(d+1)}_{i=d+1}=\om^{(d)}(x_\mu)$ and $\Om^{(d+1)}_{i=d+1}=\Om^{(d)}(x_\mu)$.
The density $\om^{(d)}$ in \re{SCSf} is the Wess-Zumino (WZ) term.

\subsubsection{The $W$ term in SCP definition \re{dif}-\re{vr1}}
In defining the SCS density \re{SCS0}-\re{SCSf} in $d$ dimensions, only the definition \re{vr2} was employed,
and not the variant \re{vr1}. It is nonetheless reasonable for the sake of being self-contained, to illustrate
the provenance of the gauge invariant term $W$ that appears in the crucial relation \re{dif} which
splits~\footnote{
It may be relevant to mention that such a definition for a gauge invariant and total divergence density
like $\vr$ in \re{vr1}-\re{vr2} can be made for a $SO(d)$ gauged system of $d$-tuplet Higgs field, though in
that case that is not necessary since the ``energy'' density of such monopoles are is bounded by the Higgs--Chern-Pontryagin
(HCP) density defined in \cite{Tchrakian:2010ar}. Such lower bounds as \re{vr1}-\re{vr2} for Higgs systems were discussed
in \cite{Tchrakian:2002ti}, where again this splitting becomes unique in the $d=2$ case only, and also the lower bound
is saturated when the usual Abelian Higgs model is chosen. In all $d\ge 3$, this splitting has some freedom and the relevant
lower bound is not saturated. Moreover these lower bounds are always higher than the HCP lower bounds~\cite{Tchrakian:2002ti}.
},
into the two terms $\pa_i\Om^{(d+1)}_i$ and $W$ leading to the two equivalent definitions \re{vr1} and \re{vr2}
for the SCP $\vr$. This splitting occurs for the $SO(2)$ SCP in two
dimensions~\cite{Schroers:1995he}, albeit in
a somewhat different approach to here. The Ref.~\cite{Schroers:1995he} was the template for extending this
construction to $d=3$ for $SO(3)$ and to $d=4$ for $SO(4)$, in \cite{Tchrakian:1997sj}, and subsequently to smaller
gauge groups in each case (see \cite{Tchrakian:2015pka} and references therein).

The term $W$ in the SCP density \re{vr1} is involved in stating Bogomol'nyi-like ``energy lower bounds'',
which is saturated only for the $SO(2)$ case in
$d+1=2$ dimensions presented in \cite{Schroers:1995he}. As well as this, it turns out that in dimensions $d+1\ge 3$
the splitting in \re{dif} is not always unique. To demonstrate these features, it helps to consider the $W$
terms for the special examples considered in Sections {\bf 3.1}, {\bf 3.2} and {\bf 4.1} below.

These examples pertain to the Type$_I$ $SO(2)$ and $SO(3)$ SCP densities \re{Omf} and \re{SCP42b3} respectively,
both in $d+1$ dimensions, and, to the Type$_{II}$  $SO(4)$ SCP in $d+1=4$, given by $\Om_i^{(d+1)}$ in \re{Om4f} below.
The $W$ terms for these examples are listed as
\bea
W&=&\frac12\,\vep_{ijk_1k_2\dots k_{d-1}} \vep^{A_1A_2\dots A_{d}}\,F_{ij}\,D_{k_1}\f^{A_1}
D_{k_2}\f^{A_2}\dots D_{k_{d-1}}\f^{A_{d-1}}\f^{A_{d}},\label{Wa}\\
W&=&\frac12\,\vep_{ijk_1k_2\dots k_{d-1}}\vep^{A_1A_2\dots A_{d-1}}\,F_{ij}^\al\f^\al
\,D_{k_1}\,\f^{A_1}\,D_{k_2}\,\f^{A_2}\dots D_{k_{d-1}}\,\f^{A_{d-1}},\label{Wb}\\
W&=&3!\,\vep_{ijkl}\vep^{abcd}\f^5\left\{\frac{1}{24}(\f^5)^2{F}_{ij}^{ab}{F}_{kl}^{cd}
+\frac12\,{F}_{ij}^{ab}{D}_{[k}\f^{c}{D}_{l]}\f^{d}\right\},\label{Wc}
\eea
where in \re{Wa} and \re{Wb} the indices $A_1,A_2,\dots$ label the ungauged components of the Skyrme scalar $\f^a$.

On close inspection, it is clear that a gauge-invariant and total-divergence term can be extracted fron \re{Wa} and
from the second term in \re{Wc}, while no such term can be isolated in \re{Wb}. It can be noted that this extraction can be
carried out for $SO(n)$ for even $n$ gauge groups, with the exception of $SO(2)$ gauging of the $O(3)$ sigma
model~\cite{Schroers:1995he}. Thus in cases such
as \re{Wa} and \re{Wc}, the total-divergence term extracted from the term $W$ appearing
in \re{vr1}, can be transferred to \re{vr2} and incorporated in $\pa_i\Om_i$, redefining the latter.

This is the freedom present in the splitting carried out in \re{dif}, which can be aginfully employed in casting the
resulting Bogomol'nyi-like inequalities in a useful form. Clearly, this choice influences also the definition of the
corresponding SCS densities, and in all examples encounered it leads also to an optimal choice for the latter.

\subsubsection{Gauge transformation of SCS}
Consider the transformation of $\hat\Om^{(d+1)}_i$, appearing in the definition of the SCP density $\vr=\pa_i\hat\Om^{(d+1)}_i$
in \re{SCP} (or \re{13}), under the infinitesimal gauge transformation $g(x_i)$

\be
\label{infOmdplus1}
\hat\Om^{(d+1)}_i\stackrel{g}\rightarrow\hat\Om^{(d+1)}_i+\de\hat\Om^{(d+1)}_i\,.
\ee

Since $\vr$ is gauge invariant, $i.e.,$ $\de\vr=0$, it follows that
\be
\label{pahatOm1}
\pa_i(\de\hat\Om^{(d+1)}_i)=0\,,
\ee
which allows to express $\de\hat\Om^{(d+1)}_i$ formally as
\be
\label{Omv}
\de\hat\Om^{(d+1)}_i=\vep_{ijk_1k_2\dots k_{d-1}}\,\pa_jV_{k_1k_2\dots k_{d-1}}\,.
\ee
where $V_{k_1k_2\dots k_{d-1}}$ is a totally antisymmetric tensor defined in terms of the connection and curvature fields,
as well as the Skyrme scalar.

From the definition of the SCS density \re{SCS0}, \re{Omv} implies the following transformation
\be
\label{deOmSCS}
\de\Om^{(d)}_{\rm SCS}=\de\hat\Om^{(d+1)}_{i=d+1}=
\vep_{(d+1)\mu\nu_1\nu_2\dots\nu_{d-1}}\,\pa_\mu V_{\nu_1\nu_2\dots\nu_{d-1}}
\ee
which is clearly defined on the space with the $d$-dimensional coordinates $x_\mu$.

It follows from \re{deOmSCS} that under an infinitesimal gauge transformation $g(x_\mu)$, the SCS density
$\Om^{(d)}_{\rm SCS}$ transforms as
\be
\label{infOmSCS}
\Om^{(d)}_{\rm SCS}\stackrel{g}\rightarrow
\Om^{(d)}_{\rm SCS}+\vep_{\mu\nu_1\nu_2\dots\nu_{d-1}}\,\pa_\mu V_{\nu_1\nu_2\dots\nu_{d-1}}\,,
\ee
meaning that the SCS density is {\it gauge invariant up to a total divergence,} as a result, the
Euler-Lagrange equations are {\it gauge covariant}.

\section{Type$_{I}$ $SO(2)$ and $SO(3)$ SCS in $d$ dimensions }
The largest gauge group of a SCS density in $d$ dimensions is $SO(d+1)$, namely the gauge group of the
SCP density \re{SCP} of the $O(d+2)$ Skyrme system in $d+1$ dimensions. Type$_{I}$ SCS are those which
are gauged with $SO(N)$, with $N<d+1$. What is distinctive with the Type$_{I}$ SCS is that they can be
expressed in a uniform format for a given $N$ in any dimension $d$.
This format is typified by the linear dependence of $\Om^{(d)}$ in the definition  of the SCS \re{SCSf},
on the gauge connection $A_\mu$, and the absence there of the gauge curvature $F_{\mu\nu}$.

In the next two Subsections, only the $SO(2)$ and $SO(3)$ cases are analysed, restricting attention
to one Abelian and one non-Abelian case.

\subsection{$SO(2)$ gauged SCS$_{I}$ in $d$ dimensions}
Consider the $O(d+2)$ sigma model in $d+1$ dimensions, gauged with $SO(2)$. As per the prescription 
described above, we start with
the SCP density in $d+1$ dimensions, and after the descent by one step arrive at the SCS density in $d$ dimensions.
Here only two components of the $d+2$ compnenmt $O(d+2)$ Skyrme scalar $\f^a=(\f^{\al},\f^A)$ are gauged with $SO(2)$
according to the gauging prescription
\bea
D_i\f^{\al}&=&\pa_i\f^{\al}+A_i(\vep\f)^{\al}\ ,\quad\al=1,2\ ,\quad(\vep\f)^{\al}=\vep^{\al\bt} \f^\bt \label{coval22}\\
D_i\f^{A}&=&\pa_i\f^A\label{covaA22}\ ,\qquad\qquad
A=1,2,\dots d\ ,\ {\rm or}\ \ A=3,4,\dots d+2\,.
\eea

Examples of $SO(2)$ gauged SCP densities $\pa_i\hat\Om^{(d+1)}_i$, \re{defhatOm}, in various dimensions
are listed in Ref.~\cite{Tchrakian:2015pka}, from which follows the SCP in $d+1$ dimensions 
\bea
\pa_i\hat\Om^{(d+1)}_i&=&\pa_i\,(\om^{(d+1)}_i+\Om^{(d+1)}_i)\nonumber\\
&=&\pa_i\,[\om^{(d+1)}_i+\vep_{ijk_1k_2\dots k_{d-1}} \vep^{A_1A_2\dots A_{d}}\,A_j\,\pa_{k_1}\f^{A_1}
\pa_{k_2}\f^{A_2}\dots\pa_{k_{d-1}}\f^{A_{d-1}}\f^{A_{d}}]\label{Omf}\,,
\eea
by induction.

From \re{Omf} follows via the one-step descent \re{SCP}-\re{SCS1} or \re{defhatOm}-\re{SCSf}, fixing $i=d+1$,
the SCS density \re{SCSf} in $d$ dimensions
\[
\Om_{\rm SCS}^{(d)}=\om^{(d)}+\Om^{(d)}
\]
in which $\Om^{(d)}$ is given by
\bea
\label{Omd}
\Om^{(d)}
&=&\vep_{\nu\mu_1\mu_2\dots\mu_{d-1}} \vep^{A_1A_2\dots A_{d}}\,A_{{\nu}}\,\pa_{\mu_1}\f^{A_1}
\pa_{\mu_2}\f^{A_2}\dots\pa_{\mu_{d-1}}\f^{A_{d-1}}\f^{A_{d}}\,.
\eea

To evaluate the WZ term $\om^{(d)}$ in \re{SCSf} one must first evaluate $\vr_0$ defined by \re{wnd} in $d+1$
dimensions, cast it in total divergence form $\pa_i\om^{(d+1)}$, and then perform the
one-step descent \re{SCP}-\re{SCS1} or \re{defhatOm}-\re{SCSf}, fixing $i=d+1$. For this, it is
necessary to employ a parametrisation which is compliant with
the constraint $|\f^a|^2=(|\f^\al|^2+|\f^A|^2)=1$,
\be
\label{para22}
\f^{\al}=\sin f\,n^{\al}\ ,\quad\f^A=\cos f\,n^A\ ;\ \al=1,2;\ A=1,2,\dots,d\,,
\ee
where $n^\al$ and $n^A$ are vector valued functions of $unit\ length$. The two-component unit vector
$n^\al=(\cos\psi,\sin\psi)$ being parametrised by the angular
coordinate $\psi$ on $S^1$, and the $(d-1)$-component unit vector $n^A$ by the coordinates on $S^{d-2}$.

Substituting \re{para22} in \re{wnd}, and noting that $\vep^{\al\bt}n^\al\pa_{j_{d+1}}n^\bt=\pa_{j_{d+1}}\psi$
\bea
\label{vr0comp}
\vr_0&=&-\,\vep_{ijk_1k_2\dots k_{d-1}} (\pa_i\cos^df)\,(\vep^{\al\bt}n^\al\pa_{j}n^\bt)\,
(\vep^{A_1A_2\dots A_{d}}\,\pa_{k_1}n^{A_1}
\pa_{k_2}n^{A_2}\dots\pa_{k_{d-1}}n^{A_{d-1}}n^{A_{d}})\nonumber\\
&=&-\,\pa_i\left[(\cos^df)\,\vep_{ij}\,(\pa_j\psi)\,
(\vep_{k_1k_2\dots k_{d-1}}\vep^{A_1A_2\dots A_{d}}\,\pa_{k_1}n^{A_1}
\pa_{k_2}n^{A_2}\dots\pa_{k_{d-1}}n^{A_{d-1}}n^{A_{d}})\right]
\eea
from which follows the WZ term
\be
\label{WZ}
\om^{(d)}=-\vep_{\nu\mu_1\mu_2\dots\mu_{d-1}}\,\cos^df\, (\vep^{A_1A_2\dots A_{d}}\,
\pa_{\mu_1}n^{A_1}\pa_{\mu_2}n^{A_2}\dots\pa_{\mu_{d-1}}n^{A_{d-1}}n^{A_{d}})\,\pa_\nu\psi\,.
\ee

Next, we evaluate $\Om^{(d)}$ in the parametrisation \re{para22} by
substituting the latter in \re{Omd}, yielding
\be
\label{Omdp}
\Om^{(d)}=\vep_{\nu\mu_1\mu_2\dots\mu_{d-1}}\,\cos^df
\,(\vep^{A_1A_2\dots A_{d}}\,\pa_{\mu_1}n^{A_1}
\pa_{\mu_2}n^{A_2}\dots\pa_{\mu_{d-1}}n^{A_{d-1}}n^{A_{d}})\,A_\nu
\,,
\ee
and adding \re{WZ} to \re{Omdp} we end up with the SCS density \re{SCSf} in $d$ dimensions
\be
\label{SCSfin}
\Om^{(d)}_{\rm SCS}=\vep_{\nu\mu_1\mu_2\dots\mu_{d-1}}\,\cos^df
\,(\vep^{A_1A_2\dots A_{d}}\,\pa_{\mu_1}n^{A_1}
\pa_{\mu_2}n^{A_2}\dots\pa_{\mu_{d-1}}n^{A_{d-1}}n^{A_{d}})\,(A_\nu-\pa_\nu\psi)
\,.
\ee
Since according to the gauging prescription \re{coval22}-\re{covaA22} the scalar function $f$ and
the vector function $n^A$ are inert under gauge transformations, it follows that the SCS density
\re{SCSfin} is invariant under the Abelian gauge transformation
\be
\label{Abel}
A_\nu\to A_\nu+\pa_\nu\La
\ee
with $\psi$ in \re{SCSfin} compensating for $\La$ in \re{Abel}.

The remarkable feature here is the fact that the Abelian SCS action \re{SCSfin} is {\it explicitly gauge invariant},
unlike the usual Chern-Simons term which is seen in Subsection {\bf 2.1.1} to be
{\it gauge invariant up to total divergence} only. It appears that in this case,
the gauge transformation of the Abelian connection is compensated completely by that of the Skyrme scalar.

\subsection{$SO(3)$ gauged SCS$_{I}$ in $d$ dimensions}
The gauging prescription for the $O(d+2)$ sigma model in $d+1$ dimensions is given by
the definition of the covariant derivatives
\bea
D_i\f^{\al}&=&\pa_i\f^{\al}+A_i^{\al\bt}\,\f^\bt\ ,\quad\al=1,2,3\label{coval23}\\
D_i\f^{A}&=&\pa_i\f^A\label{covaA23}\ ,\qquad\qquad A=1,2,\dots,d-1\ ,\ \ {\rm or}\ \ A=4,5,\dots,d+2
\eea
where in \re{coval23}, $A_i^{\al\bt}=\vep^{\al\bt\ga}A_i^\ga$.

Examples of $SO(3)$ gauged SCP densities $\pa_i\hat\Om^{(d+1)}_i$, \re{defhatOm}, in various dimensions
are listed in Ref.~\cite{Tchrakian:2015pka}, from which follows the SCP in $d+1$ dimensions
\bea
\pa_i\hat\Om^{(d+1)}_i
&=&\pa_i(\om^{(d+1)}_i+\Om^{(d+1)}_i)\label{scp42a3}\\
&=&\pa_i\big(\om_i^{(d+1)}+
\vep_{ijk_1k_2\dots k_{d-1}}\,A_j^\al\f^\al\cdot\nonumber\\
&&\qquad\qquad\cdot\,
\vep^{A_1A_2\dots A_{d-1}}\,\pa_{k_1}\,\f^{A_1}\,\pa_{k_2}\,\f^{A_2}\dots\pa_{k_{d-1}}\,\f^{A_{d-1}}\big)\label{SCP42b3}
\eea
by induction.

From \re{SCP42b3} follows via the one-step descent \re{SCP}-\re{SCS1} or \re{defhatOm}-\re{SCSf}, fixing $i=d+1$,
the SCS density \re{SCSf} in $d$ dimensions
\[
\Om_{\rm SCS}^{(d)}=\om^{(d)}+\Om^{(d)}
\]
in which $\Om^{(d)}$ is given by
\be
\label{Omd3}
\Om^{(d)}=\vep_{\nu\mu_1\mu_2\dots\mu_{d-1}}
\vep^{A_1A_2\dots A_{d-1}}\,A_\nu^\al\f^\al\,
\pa_{\mu_1}\,\f^{A_1}\,\pa_{\mu_2}\,\f^{A_2}\dots\pa_{\mu_{d-1}}\,\f^{A_{d-1}}
\ee

As in the previous case with $SO(2)$ gauging, to evaluate the WZ term $\om^{(d)}$ in \re{SCSf} we must
evaluate the winding number density $\vr_0$ defined by \re{wnd} in $d+1$ and
cast it in total divergence form $\pa_i\om^{(d+1)}$. After that the
one-step descent \re{SCP}-\re{SCS1} (or \re{defhatOm}-\re{SCSf}) can be performed, fixing $i=d+1$.
As was done in \re{para22} above, this is achieved by employing a constraint compliant parametrisation
which in this case is formally that as \re{para22}, namely
\be
\label{para3}
\f^{\al}=\sin f\,n^{\al}\ ,\quad\f^A=\cos f\,n^A\ ;\ \al=1,2,3;\ A=1,2,\dots,d-1\,,
\ee
with $n^\al$ and $n^A$ being $2$ and $(d-1)$ component $unit$ vectors respectively, $n^\al=n^\al(\chi,\psi)$ being
parametrised by the polar and azimuthal coordinate on $S^2$, and $n^A$ by the angular coordinates on $S^{d-2}$.

Substituting \re{para3} in \re{wnd} yields,
\bea
\label{vr0comp3}
\vr_0&=&\frac14\,\vep_{ijkl_1l_2\dots l_{d-2}}(\pa_i\F^{(d)})\,
(\vep^{\al\bt\ga}n^\al\pa_{j}n^\bt\pa_k n^\ga)\, (\vep^{A_1A_2\dots A_{d-1}}\,\pa_{l_1}n^{A_1}
\pa_{l_2}n^{A_2}\dots\pa_{l_{d-2}}n^{A_{d-2}}n^{A_{d-1}})\nonumber\\
&=&\frac14\,\pa_i\big[\F^{(d)}\,\vep_{ijkl_1l_2\dots l_{d-2}}
(\vep^{\al\bt\ga}n^\al\pa_{j}n^\bt\pa_k n^\ga)\cdot\nonumber\\
&&\qquad\qquad\qquad\qquad\qquad\cdot (\vep^{A_1A_2\dots A_{d-1}}\,\pa_{l_1}n^{A_1}
\pa_{l_2}n^{A_2}\dots\pa_{l_{d-2}}n^{A_{d-2}}n^{A_{d-1}})\big]
\eea
in which the symbol $\F^{(d)}$ is
\bea
\F^{(d)}&=&\frac{1}{d+1}\,\left(f-\frac{1}{d+1}\sin(d+1)f\right)\ ,\quad{\rm for\ odd}\ d\,,\label{Fevend}\\
&=&\sin^{d+1}f\ ,\qquad\qquad\qquad\qquad\qquad\ \ \ {\rm for\ even}\ d\,.\label{Foddd}
\eea

It follows from \re{vr0comp3} that the WZ term $\om^{(d)}$ is
\bea
\label{WZ3}
\om^{(d)}&=&\frac14\,\F^{(d)}\vep_{\nu_1\nu_2\mu_1\mu_2\dots \mu_{d-2}}
(\vep^{\al\bt\ga}n^\al\pa_{\nu_1}n^\bt\pa_{\nu_2} n^\ga)\cdot\nonumber\\
&&\qquad\qquad\qquad\qquad\qquad\cdot (\vep^{A_1A_2\dots A_{d-1}}\,\pa_{\mu_1}n^{A_1}
\pa_{\mu_2}n^{A_2}\dots\pa_{\mu_{d-2}}n^{A_{d-2}}n^{A_{d-1}})\,.
\eea

We next evaluate $\Om^{(d)}$ given by \re{Omd3} in the parametrisation \re{para3}
\bea
\label{Om3}
\Om^{(d)}&=&-\vep_{\nu_1\nu_2\mu_1\mu_2\dots \mu_{d-2}}
(A_{\nu_1}^\al\,n^\al\ \pa_{\nu_2}\F^{(d)})\cdot\nonumber\\
&&\qquad\qquad\qquad\qquad\qquad\cdot (\vep^{A_1A_2\dots A_{d-1}}\,\pa_{\mu_1}n^{A_1}
\pa_{\mu_2}n^{A_2}\dots\pa_{\mu_{d-2}}n^{A_{d-2}}n^{A_{d-1}})\,.
\eea
and finally adding \re{WZ3} and \re{Om3}, we have the SCS density
\bea
\label{SCS3}
\Om^{(d)}_{\rm SCS}&=&\frac14\,\vep_{\nu_1\nu_2\mu_1\mu_2\dots \mu_{d-2}}
(\vep^{A_1A_2\dots A_{d-1}}\,\pa_{\mu_1}n^{A_1}
\pa_{\mu_2}n^{A_2}\dots\pa_{\mu_{d-2}}n^{A_{d-3}}n^{A_{d-2}})\cdot
\nonumber\\
&&\qquad\qquad\qquad\qquad\qquad\cdot
(4\,\pa_{\nu_1}\F^{(d)}A_{\nu_2}^\al\,n^\al
+\F^{(d)}\,\vep^{\al\bt\ga}n^\al\pa_{\nu_1}n^\bt\pa_{\nu_2} n^\ga)\,.
\eea

\subsubsection{Gauge dependence}
It is useful to express \re{SCS3} as
\bea
\label{SCS3x}
\Om^{(d)}_{\rm SCS}&=&\frac14\,\Xi_{\nu_1\nu_2}\,
(4\,\pa_{\nu_1}\F^{(d)}A_{\nu_2}^\al\,n^\al
+\F^{(d)}\,\vep^{\al\bt\ga}n^\al\pa_{\nu_1}n^\bt\pa_{\nu_2} n^\ga)\,.
\eea
where the prefactor
\be
\label{prefactor}
\Xi_{\nu_1\nu_2}=\vep_{\nu_1\nu_2\mu_1\mu_2\dots \mu_{d-2}}
\vep^{A_1A_2\dots A_{d-1}}\,\pa_{\mu_1}n^{A_1}
\pa_{\mu_2}n^{A_2}\dots\pa_{\mu_{d-2}}n^{A_{d-2}}n^{A_{d-1}}
\ee
is gauge invariant.

Instead of working with the real parametrisation of the $SO(3)$ gauge group, it is convenient to work with
the $SU(2)$ parametrisation
\be
\label{SU2}
g=\cos\frac{\La}{2}+i\,\vec\si\cdot\vec m\,\sin\frac{\La}{2}\ ,\quad {\rm with}|\quad\vec m|^2=1
\ee
such that the vector $n^\al$ and the connection $A_\mu^\al$ are now expressed as $SU(2)$ algebra elements
\be
\label{not}
n=n^\al\,\si^\al\equiv\vec n\cdot\vec\si\ ,\quad A_\mu=A_\mu^\al\,\si^\al\equiv\vec A_\mu\cdot\vec\si
\ee
and they transform under the $SU(2)$ gauge gauge group element $g$ as
\bea
n&\stackrel{g}\to& g^{-1}\,n\, g\\
A_\mu&\stackrel{g}\to& g^{-1}\,A_\mu g+i\,g^{-1}\,\pa_\mu\ g\,,
\eea

In the parametrisation \re{not} the action \re{SCS3x} is expressed as
\be
\label{SCSgenTr}
\Om^{(d)}_{\rm SCS}=\frac12\,\Xi_{\mu\nu}\,\left\{4\,\pa_\mu\F^{(d)}\,
\mbox{Tr}(n\,A_\nu) -i\,\F^{(d)}\,\mbox{Tr}(n\ \pa_\mu n\ \pa_\nu n)\right\}\,.
\ee

Under infinitesimal transformation with $\cos\La\approx \eins$ and $\sin\La\approx\La$ in \re{SU2}, $g$
reduces to
\be
\label{SU2inf}
g=\eins+\frac i2\,\La\,\vec m\cdot\vec\si
\ee
under which $n$ transforms, up to first order in $\La$, as
\be
\label{gaugen}
n=\vec n\cdot\vec\si\stackrel{g}\to \vec n\cdot\vec\si-\vec n\times(\La \vec m)\cdot\vec\si\,,
\ee
and hence the term $\mbox{Tr}(n\,A_\nu)$ in \re{SCSgenTr} as
\bea
\mbox{Tr}(n\,A_\nu)&\stackrel{g}\to&\mbox{Tr}(n\,A_\nu)+i\,\mbox{Tr}(n\,\pa_\nu g\,g^{-1})\label{1}\\
&=&\mbox{Tr}(n\,A_\nu)-\vec n\cdot\pa_\nu(\La\,\vec m)\label{2}\\
&=&2\,\vec{n}\cdot\vec{A_\mu}-\vec{n}\cdot\pa_\mu(\La\vec m)\,,\label{3}
\eea
and the term $\Xi_{\mu\nu}\mbox{Tr}(n\ \pa_\mu n\ \pa_\nu n)$ in \re{SCSgenTr}, as
\bea
\Xi_{\mu\nu}\mbox{Tr}(n\ \pa_\mu n\ \pa_\nu n)&\stackrel{g}\to&\vep_{\mu\nu\dots}\mbox{Tr}(g^{-1}n\,g)\,
\pa_\mu(g^{-1}n\,g)\,\pa_\nu(g^{-1}n\,g)\label{1a}\\
&=&2i\,\vep_{\mu\nu\dots}\left[\vec{n}\cdot(\pa_\mu\vec n\times\pa_\nu\vec n)
-2\,\pa_\mu\vec n\cdot\pa_\nu(\La\vec m)\label{2a}
\right]\,.
\eea

The result is
\bea
\Om_{\rm SCS}^{(d)}&\stackrel{g}\to&\Om_{\rm SCS}^{(d)}-2\,\Xi_{\mu\nu}
\left[\pa_\mu\F^{(d)}\,\vec n\cdot\pa_\nu(\La\vec m)+\F^{(d)}\,\pa_\mu\vec n\cdot\pa_\nu(\La\vec m)\right]\nonumber\\
&=&\Om_{\rm SCS}^{(d)}-2\,\Xi_{\mu\nu}\,\pa_\mu\left[\F^{(d)}\,\vec n\cdot\pa_\nu(\La\vec m)\right]\,,
\label{SCSgen1}
\eea
and since $\Xi_{\mu\nu}$ is by definition \re{prefactor} antisymmetric in $\mu\nu$,
\be
\label{SCSgen2}
\Om_{\rm SCS}^{(d)}\stackrel{g}\to\Om_{\rm SCS}^{(d)}
-2\,\pa_\mu\left[\Xi_{\mu\nu}\,\F^{(d)}\,\vec n\cdot\pa_\nu(\La\vec m)\right]\,,
\ee
$i.e.,$ that the SCS density is gauge invariant up to a total divergence term
like the usual Chern-Simons density which is shown in Subsection {\bf 2.1.1}

\section{Type$_{II}$ SCS in $d=3$ dimensions}
Type$_{II}$ SCS terms are the germane analogues of the Chern-Simons (CS) densities,
in that unlike Type$_{I}$ SCS they feature both the gauge curvature $F_{\mu\nu}$ and the gauge connection $A_\mu$.
Unlike their Type$_{I}$ counterparts however, the construction of Type$_{II}$ SCS densities does not lend itself
to simple uniform formats in all dimensions $d$, and, they are defined only for non-Abelian gauge groups. Indeed,
in \cite{Tchrakian:2015pka}, which is our source for SCS densities, the only example given is that of the
$SO(4)$ gauged SCS in $d=3$ dimensions.

What was not presented in Ref.~\cite{Tchrakian:2015pka} is a Type$_{II}$ SCS in $d=3$, pertaining to gauging with
a subgroup~\footnote{
Clearly this does not include the group contraction $SO(4)\to SO(3)$, since that results in
the Type$_{I}$ SCS analysed in {\bf Section 3.2} above, for $d=3$.} of $SO(4)$. Below in {\bf Section 4.1}
is presented the Type$_{II}$ $SO(4)$ SCS in $d=3$, and the Type$_{II}$ $SO(2)\times SO(2)$ SCS in $d=3$ {\bf Section 4.2}.

\subsection{$SO(4)$ gauged Type$_{II}$ SCS in $d=3$}
This example is given in \cite{Tchrakian:2015pka} but it is repeated here, to make the presentation
of the $SO(2)\times SO(2)$ SCS in {\bf Section 4.2} self-contained.
The gauge transformation property of the Type$_{II}$ $SO(4)$ SCS is not carried out here,
since that analysis is very cumbersome.

The covariant derivative of the $O(5)$ Skyrme scalar $(\f^a,\f^5)\ ,\ a=1,2,3,4$ is
\bea
D_i\f^{a}&=&\pa_i\f^{a}+A_i^{ab}\,\f^b\ ,\quad a=1,2,3,4\label{coval234}\\
D_i\f^{5}&=&\pa_i\f^5\label{cov5}\,.
\eea
To define the SCS density, what is needed are the definitions of $\om^{(3+1)}$ and $\Om^{(3+1)}$ in
\re{defhatOm}, from which follows the definition of the SCS density by \re{SCS1}. The quantity
$\hat\Om^{(3+1)}$ in \re{defhatOm} in this case is,
\[
\hat\Om^{(3+1)}_i=\om^{(3+1)}_i+\Om^{(3+1)}_i\,,
\]
where the quantity $\Om_i^{(3+1)}$ in $3+1$ dimensions, apearing in
\cite{Tchrakian:2015pka,Navarro-Lerida:2016omj,Navarro-Lerida:2020jft}, is
\bea
\Om_i^{(3+1)}&=&3!\,\vep_{ijkl}\vep^{abcd}\f^5\bigg\{\frac12{F}_{kl}^{{c{d}}}\f^{a}{D}_j\f^{b}
+\pa_j\left[{A}_l^{ab}\f^{c}\left(\pa_k\f^{d}+\frac12{A}_k\f^{d}\right)\right]+\nonumber\\
&&\qquad\qquad\qquad\ \ \ \ +\frac14\left(1-\frac13(\f^5)^2\right){A}_l^{ab}\left[\pa_j{A}_k^{cd}+\frac23({A}_j{A}_k)^{cd}\right]\bigg\} \ ,  \label{Om4f}
\eea
from which follows by the one-step descent, the quantity $\Om^{(3)}$ in \re{SCSf}
\bea
\Om^{(3)}&=&3!\,\vep_{\mu\nu\la}\vep^{abcd}\f^5\bigg\{-\frac12{F}_{\mu\nu}^{{c{d}}}\f^{a}{D}_\la\f^{b}
+\pa_\mu\left[{A}_\la^{ab}\f^{c}\left(\pa_\nu\f^{d}+\frac12{A}_\nu\f^{d}\right)\right]+\nonumber\\
&&\qquad\qquad\qquad\ \ \ \ +\frac14\left(1-\frac13(\f^5)^2\right){A}_\la^{ab}\left[\pa_\mu{A}_\nu^{cd}+\frac23({A}_\mu{A}_\nu)^{cd}\right]\bigg\} \,.  \label{Om4ff}
\eea
This density is clearly {\it gauge variant}, seen for example from the
term multiplying $\left(1-\frac13(\f^5)^2\right)$ in the second line, which is the (Euler,
rather than the Pontryagin) Chern-Simons density for the $SO(4)$ gauge field in $d=3$.


To evaluate the term $\om^{(3)}$ in \re{SCSf} however, it is necessary to employ the
constraint compliant parametrisation
\be
\label{gcp4}
\f^a=\sin f\,n^a\,,\quad\f^5=\cos f\,,\quad |n^a|^2=1
\ee
of the Skyrme scalar $(\f^a,\f^5)$.

To this end, one evaluates $\vr_0$ in \re{vr2} defined by \re{wnd}, yielding
\bea
\label{vr04}
\vr_0&=&-4\,\vep_{ijkl}\vep^{abcd}\,\pa_i\left(\f^5-\frac13(\f^5)^3\right)\cdot
n^a\pa_jn^b\,\pa_kn^c\,\pa_ln^d\nonumber\\
&=&-4\,\vep_{ijkl}\vep^{abcd}\,\pa_i\left[\left(\f^5-\frac13(\f^5)^3\right)\,
n^a\pa_jn^b\,\pa_kn^c\,\pa_ln^d\right]\nonumber\\
&\stackrel{\rm def.}=&\pa_i\,\om_i^{(3+1)}\label{wnd4}
\eea
whence we have
\be
\label{om34}
\om^{(3)}\stackrel{\rm def.}=\om_{i=4}^{(3+1)}=
4\,\vep_{\mu\nu\la}\vep^{abcd}\,\left(\f^5-\frac13(\f^5)^3\right)\,
n^a\pa_\mu n^b\,\pa_\nu n^c\,\pa_\la n^d\,.
\ee

Adding \re{om34} and \re{Om4ff}
\bea
\label{SCSf4}
\Om^{(3)}_{\rm SCS}
&=&\om^{(3)}+\Om^{(3)}
\eea
for $SO(4)$ gauging.

The WZ term $\om^{(3)}$ in \re{SCSf4} is given in the constraint-compliant parametrisation
\re{gcp4}, but this is not done for the term $\Om^{(3)}$ here.

\subsection{$SO(2)\times SO(2)$ gauged Type$_{II}$ SCS in $d=3$}
The task is as above the calculation of $\Om^{(3)}_{\rm SCS}=\om^{(3)}+\Om^{(3)}$, \re{SCSf}. It is useful to start with
the calculation of $\Om^{(3)}$ for this case by subjecting \re{Om4ff} to the gauge group contraction
\[
SO(4)\to SO(2)\times SO(2)
\]
on the gauge connection $A_\mu^{ab}=(A_\mu^{\al\bt},A_\mu^{AB},A_\mu^{\al A})$
\be
\label{contr}
A_\mu^{\al\bt}=A_\mu\,\vep^{\al\bt}\ ,\ \ A_\mu^{AB}=B_\mu\,\vep^{AB}\ , \ \ A_\mu^{\al A}=0\,,
\ee
where $A_\mu$ and $B_\mu$ are the connections of the two (distinct) Abelian subgroups of $SO(4)$.

We denote the corresponding components of the Abelian curvature by
\be
\label{contrcurv}
F_{\mu\nu}^{\al\bt}=F_{\mu\nu}\,\vep^{\al\bt}\ ,\ \ F_{\mu\nu}^{AB}=G_{\mu\nu}\,\vep^{AB}\ ,\ \  F_{\mu\nu}^{\al A}=0\,,
\ee
where
\[
F_{\mu\nu}=\pa_{\mu}A_{\nu}-\pa_{\nu}A_{\mu}\ \ {\rm and}\ \ G_{\mu\nu}=\pa_{\mu}B_{\nu}-\pa_{\nu}B_{\mu}
\]
in an obvious notation.

The corresponding $SO(4)$ cavariant derivatives contract to
\bea
D_\mu\f^{\al}&=&\pa_\mu\f^{\al}+A_\mu\,(\vep\f)^{\al}\ ,\quad \al=1,2\ ,\quad(\vep\f)^{\al}
=\vep^{\al\bt} \f^\bt\label{coval222}\\
D_\mu\f^{A}&=&\pa_\mu\f^{A}+B_\mu\,(\vep\f)^A\ ,\quad A=3,4\ ,\quad(\vep\f)^A=\vep^{AB}\f^B\label{covA22}\\
D_\mu\f^{5}&=&\pa_\mu\f^5\label{cov522}\,.
\eea

Substituting \re{contr}, \re{contrcurv} and \re{coval22}-\re{cov522} in \re{Om4ff}, we obtain
$\Om^{(3)}$ for the gauge group $SO(2)\times SO(2)$
\bea
\label{SCSAB}
\Om^{(3)}&=&3!\,\vep^{4\mu\nu\la}\ \f^5\bigg\{\frac13\,(\f^5)^2
(A_\la G_{\mu\nu}+B_\la F_{\mu\nu})\nonumber\\
&&\qquad\qquad\quad-A_\mu B_\nu\ \pa_\la(|\f^\al|^2-|\f^A|^2)\nonumber\\
&&\qquad\qquad\quad 
-2\left[A_\la\ (\vep\pa_\mu\f)^A\,\pa_\nu\f^A+B_\la\ (\vep\pa_\mu\f)^\al\,\pa_\nu\f^\al\right]\bigg\}\,.
\eea

Next, calculate $\om^{(3)}$ in \re{SCSf} to complete the construction of the SCS density $\Om^{(3)}_{\rm SCS}$.
This can be done only after expressing $\vr_0=\pa_i\om_i^{(3+1)}$ in {\it constraint compliant} parametrisation
of the $O(5)$ Skyrme scalar
$(\f^a,\f^5)$, with $\f^a=(\f^\al,\f^A)\ ,\ \ \al=1,2\ ;\ A=3,4$, which is
\be
\label{f}
\f^\al=\sin f\sin g\ n_{(1)}^{\al}\ ,\ \ \f^A=\sin f\cos g\ n_{(2)}^{A}\ ,\ \ \f^5=\cos f
\ee
\be
\label{n5}
n_{(1)}^{\al}=\left(\begin{array}{l}
\cos\psi\\
\sin\psi\\
\end{array}\right)\ ,\quad
n_{(2)}^{A}=\left(\begin{array}{l}
\cos\chi\\
\sin\chi\\
\end{array}\right)\,.
\ee
The result is
\bea
\vr_0&=&-3!\cdot 2\,\vep_{ijkl}\,\left[\pa_i\left(\cos f
-\frac13\cos^3f \right)\right]\,(\pa_j\sin^2g)\,\pa_k\psi\,\pa_l\chi\label{vr01},
\eea
from which one can choose $e.g.,$ to extract the partial derivative $\pa_i$, resulting in
$\om^{(3)}\stackrel{\rm def.}=\om_{i=4}^{(3+1)}$
\bea
\om^{(3)}&=&
-3!\cdot 2\,\vep^{4\la\mu\nu}\left(\cos f-\frac13\cos^3f
\right)\,(\pa_\la\sin^2g)\,\pa_\mu\psi\,\pa_\nu\chi.
\label{WZn}
\eea

Finally, the SCS density for the gauge group $SO(2)\times SO(2)$
\[
\Om^{(3)}_{\rm SCS}=\om^{(3)}+\Om^{(3)}.
\]
expressed entirely in the parametrisation \re{f}-\re{n5}, is
  
\bea
\label{SCSgaugecomp}
\frac{1}{3!}\Om^{(3)}_{\rm SCS}&=&\vep^{\mu\nu\la}\bigg\{
-2\,\left(\cos f-\frac13\cos^3f
\right)\,(\pa_\la\sin^2g)\,\pa_\mu\psi\,\pa_\nu\chi\nonumber\\
&&\qquad\qquad\quad+\frac13\,\cos^3f\,(A_\la G_{\mu\nu}+B_\la F_{\mu\nu})\nonumber\\
&&\qquad\qquad\quad-A_\mu B_\nu\,\cos f\left[\pa_\la(\sin^2f\sin^2g)-\pa_\la(\sin^2f\cos^2g)\right]\nonumber\\
&&\qquad\qquad\quad 
+2\left[A_\la\,\pa_\mu(\sin^2f\sin^2g)\ \pa_\nu\chi+B_\la\,\pa_\mu(\sin^2f\cos^2g)\ \pa_\nu\psi\right]\bigg\}.
\eea
Setting $B_\mu=0$ in \re{SCSgaugecomp} results in the Type$_I$ SCS density \re{SCSfin} for $d=3$.

The action \re{SCSgaugecomp} is manifestly {\it gauge variant}. The question is, as a Chern-Simons like density, is it
like the latter {\it gauge invariant up to total divergence}? A straightforward way to test this is to show
that the resulting Euler-Lagrange equations are gauge covariant.

The variational equations w.r.t. $A_\tau$ $resp.$ $B_\tau$ are
\bea
\vep^{\tau\mu\nu}\,\cos f\left[\frac13\,\cos^2 f\,G_{\mu\nu}-\pa_\mu(\sin^2f\cos^2g)\,(B_\nu-\pa_\nu\chi)\right]
&=&0,\label{varA}\\
\vep^{\tau\mu\nu}\,\cos f\left[\frac13\,\cos^2 f\,F_{\mu\nu}-\pa_\mu(\sin^2f\sin^2g)\,(A_\nu-\pa_\nu\psi)\right]
&=&0,\label{varB}
\eea
and the variational equations w.r.t. $\psi$ $resp.$ $\chi$ are
\bea
\vep^{\tau\mu\nu}\left[\frac12\,G_{\mu\nu}\,\cos f\,\pa_\tau(\sin^2f\sin^2g)
+\pa_\mu f\sin^3f\,\pa_\tau(\sin^2g)\,(B_\nu-\pa_\nu\chi)\right]&=&0,\label{varpsi}\\
\vep^{\tau\mu\nu}\left[\frac12\,F_{\mu\nu}\,\cos f\,\pa_\tau(\sin^2f\cos^2g)
-\pa_\mu f\sin^3f\,\pa_\tau(\cos^2g)\,(A_\nu-\pa_\nu\psi)\right]&=&0,\label{varchi}
\eea
which are manifestly gauge invariant. The variational equations w.r.t. $f$ and $g$ are
also gauge invariant but they are too cumbersome to display here.

It may be relevant to comment on the degenerate model resulting from replacing $B_\mu=A_\mu$ in \re{SCSAB}. That
model features only the Abelian field $(A_\mu,F_{\mu\nu})$. The variational equation w.r.t. $A_\mu$ will be gauge invariant
only if one sets the functions $\psi=\chi$, which is absurd as these angles parametrise the components $\f^\al$
and $\f^A$ of the $O(5)$ Skyrme scalar, which are independent degrees of freedom, $i.e.,$ that the submodel of
\re{SCSgaugecomp} with $B_\mu=A_\mu$ leads to {\it gauge variant} equations of motion.

\section{Summary and outlook}
In this note, aspects of the Skyrme--Chern-Simons (SCS) densities proposed in
Ref.~\cite{Tchrakian:2015pka} are elaborated on,
with the main emphasis being on the {\it gauge dependance} of the SCS~\footnote{
This question does not arise in the case of the Higgs--Chern-Simons (HCS) densities also proposed in \cite{Tchrakian:2015pka},
since the HCS densities turn out to be {\it gauge invariant} in even dimensions, and in odd dimensions they consist of
a gauge invariant part, plus the usual (Euler) Chern-Simons density.
}.
It is shown that the SCS actions in $d$ spacetime dimensions, as in the case of the usual Chern-Simons (CS) densities,
are {\it gauge invariant up to total divergence} and hence that their
Euler-Lagrange equations are {\it gauge invariant}.
The importance of verifying  gauge invariance of the equations of any given model is,
that in the concrete application of such actions, this is a necessary check of the correctness of the model at hand.

In Section {\bf 3}, Type$_{I}$ SCS in $d$ dimensions are presented. In Subsection {\bf 3.1}
the case with gauge group $SO(2)$, and in {\bf 3.2} with gauge group $SO(3)$. respectively.
Type$_{I}$ SCS in $d$ spacetime dimensions, are those gauged with $SO(N)$
with $N<d+1$, $SO(d+1)$ being the largest gauge group allowed.

Concrete analyses are
carried out in two examples, $SO(2)$ and $SO(3)$, one Abelian and one non-Abelian.
It is found that $SO(2)$ Type$_{I}$ SCS are explicitly gauge invariant as they stand, and hence
the resulting equations of motion are automatically gauge covariant.
By contrast, $SO(3)$ Type$_{I}$ SCS is gauge invariant only up to a total divergence, and hence their equations of motion
are gauge invariant as described in Section {\bf 2.2.2}.
(One can surmise that this property for $SO(3)$ holds in all non-Abelian gauge groups as well.)
Thus, like the usual CS (in odd dimensions) Type$_{I}$ SCS (in all dimensions) lead to gauge covariant equations of motion.
A distinguishing feature of Type$_{I}$ SCS is that the density $\Om^{(d)}$ in \re{SCSf} features only
one power of the gauge connection $A_\mu$, and no curvature term $F_{\mu\nu}$. This sets a limitation on the
application of Type$_{I}$ SCS in the context of static fields, since
for static fields the term $\om^{(d)}$ in \re{SCSf}
vanishes, while the term $\Om^{(d)}$ now displays only the temporal component $A_0$ of the gauge connection. But
we know from the results of
Refs.~\cite{Navarro-Lerida:2016omj,Navarro-Lerida:2018siw,Navarro-Lerida:2018giv,Navarro-Lerida:2020hph}
that the important new features resulting from Chern-Simons dynamics hinge on the interrelation of the electric
and the magnetic fields $A_0$ and $A_i$.
Of course in odd dimensions, when the usual CS term is present, the Type$_{I}$ SCS will have a quantitative effect.

In Section {\bf 4}, Type$_{II}$ SCS in $d=3$ dimensions are presented.
Type$_{II}$ SCS are those gauged with the largest allowed gauge group $SO(d+1)$ or with some
direct product of subgroups of $SO(d+1)$. In these notes attention is restricted to dimension $d=3$, thus to the gauge
group  $SO(4)$ and and its subgroup $SO(2)\times SO(2)$. In Subsection {\bf 4.1} gauge group $SO(4)$ is considered, and in
Subsection {\bf 4.2} gauge group $SO(2)\times SO(2)$ is analysed concretely.

Type$_{II}$ SCS feature both connection and curvature $(A_\mu,F_{\mu\nu})$ so that in the static limit both
 electric and magnetic fields $(A_0,A_i)$ will persist. This is the important aspect distinguishing
Type$_{II}$ SCS from Type$_{I}$. It is the Type$_{II}$ SCS
that promise to reproduce the special effects of Chern-Simons dynamics observed in
Refs.~\cite{Navarro-Lerida:2016omj,Navarro-Lerida:2018siw,Navarro-Lerida:2018giv,Navarro-Lerida:2020hph} in odd dimensions .
In this sense, it is the Type$_{II}$ SCS that are the germane extensions of the usual Chern-Simons densities, with the added
all important feature that they are defined in both odd and even dimensions.

In the absence of a Chern-Simons action in even dimensional spacetime, the SCS action which is defined in all dimensions
is potentially important. Of special importance is the 
physical Minkowskian $3+1$ dimensional theory. In that case such an action is also the ``anomaly related
density'' appearing in Ref.~\cite{Callan:1983nx} for the
$U(1)$ gauged Skyrme theory. Specifically, the latter is akin to the SCS$_{II}$ in that it displays both
the $SO(2)$ connection $A_\mu$ and the curvature $F_{\mu\nu}$.
A significant difference between the SCS action $\Om^{(d=4)}_{\rm SCS}$
defined by \re{SCS0}-\re{SCSf} proposed here,
and the ``anomaly related term'' appearing in \cite{Callan:1983nx},
is that in the latter the the Skyrme scalar
employed in its construction is the $O(4)$ sigma model field that supports the (topologically stable) Skyrmion,
while by contrast the Skyrme scalar employed in the construction of the SCS $\Om^{(d=4)}_{\rm SCS}$
is the $O(6)$ sigma model field
which in $3+1$ dimensions does not support a soliton~\footnote{This is simply because the SCS in $3+1$
dimensions is descended, much in the same spirit proposed earlier in Ref.~\cite{Witten:1983tw},
from a SCP density in $5$ dimensions which is defined in terms of the $O(6)$ sigma model scalar.}.

Generally, the SCS can be empoyed to invesitigate
the effects of Chern-Simons--like dynamics in both even and odd dimensions. In this context, the case of $3+1$ dimensions is
special~\footnote{The action in \cite{Callan:1983nx} follows from the results derived in Ref.~\cite{Witten:1983tw},
which is specific to $4$ dimensions. We are not aware of higher dimensional versions of this result in the iterature.}
since in that case the ``anomaly related term'' of \cite{Callan:1983nx} is also a candidate for this role.
Moreover, the latter is defined in terms of the $O(4)$ Skyrme scalar unlike its SCS counterpart that is defined
in terms of the $O(6)$ Skyrme scalar.

It would be interesting to compare the potential roles of the Chern-Simons--like densities in \cite{Callan:1983nx} and the
SCS action $\Om^{(d=4)}_{\rm SCS}$ proposed here. In both cases the topologically stable Skyrmion stabilised by the
baryon number is deformed first by the Abelian gauge field, after which the respective Chern-Simons--like action
further influences the dynamics. In the first~\cite{Callan:1983nx} case no new (scalar) field is involved while in the
second~\cite{Tchrakian:2015pka} case the $O(6)$ Skyrme scalar enters.

The largest group with which the $O(6)$ scalar can be can be gauged is $SO(5)$. The $O(4)$ Skyrme scalar
describing the usual $SO(2)$ gauged Skyrmion, can interact with the $O(6)$ Skyrme scalar
describing the SCS density only through the $SO(2)$ gauge field. Thus the $O(6)$ scalar must be gauged with
the subgroup $SO(2\times SO(3))$ of $SO(5)$, in which the $SO(3)$ gauge sector can play the role of an ``auxiliary gauge
field''~\footnote{ There is also the academic possibility of starting with the $SO(3)$ gauged $O(4)$
Skyrme model  given in Ref.~\cite{Arthur:1996np}
 which is also endowed with an ``energy lower bound''. This $SO(3)$ gauged Skyrmion is then deformed further by the
 $SO(3\times SO(2))$ gauged SCS action, so that the $O(4)$ and $O(6)$ scalars see each other via the $SO(3)$
gage field. In theis case it is the $SO(2)$ field which plays the role of ``auxiliary gauge field''.}.
Cocrete investigatons of gauged Skyrmions influenced by SCS dynamics are under active consideration at present.

\bigskip
\bigskip
\noindent
{\large\bf Acknowledgements}
My thanks for invaluable support go to Valery Rubakov. I am grateful to Francisco Navarro-Lerida (F.~N.L) and
Eugen Radu for their past collaboration in this area. Thanks to E.R. for help in preparing this report, and to F.~N.L.
for raising the question which instigated the analysis
carried out here. My thanks go to the Referee of J. Phys. A, for generous and constructive comments and sugestions.

\newpage

\begin{small}

\end{small}

\end{document}